# Efficient implementation of portfolio strategies involving cryptocurrencies and VIX INDEX and Gold

Jiahao Cui, Qiushi Li, Yuezhi Peng

**Abstract**

This research mainly explores the characteristics of different strategies and whether VIX INDEX positively influences the investment portfolio in any period. Our portfolio has six significant cryptocurrencies, VIX INDEX and gold. We perform parameter estimation on all raw data and bring the two types into different investment strategies, complete them effectively according to other characteristics, and compare the results. At the same time, we make two different portfolios, one contains VIX INDEX, and one does not have VIX INDEX. We use different portfolios in different portfolio strategies and find that VIX INDEX can positively impact the investment portfolio of cryptocurrencies, no matter in the standard market or the downward market. The research shows that gold has the same attributes as VIX INDEX and should have a specific positive effect, but no comparative experiment has been done.

## 1 INTRODUCTION

Since the advent of Bitcoin, people's understanding of cryptocurrency has continued to evolve. And in recent years, cryptocurrency has gradually entered the field of vision of more and more people. More and more countries recognize the legality of cryptocurrency (for example, Elon Musk announced that people could use Bitcoin to buy Tesla). The cryptocurrencies have attracted more institutional investors in the past years, significantly different from the 2017 Bitcoin bull market dominated by retail investors. Some Wall Street banks (Goldman Sachs, JP Morgan Chase) believe that cryptocurrencies can play a role in value/wealth storage and inflation hedging in investment portfolios and hope that "digital gold" can eventually replace traditional gold.

As a new type of risky asset, cryptocurrency favors institutional investors and occupies a place in the market investment portfolios. To evaluate how cryptocurrencies affect the maximization of the portfolios, Alla Petukhina and Erin Sprunken (2021) employs multi-asset investment strategies with digital assets using 60 German stocks, commodities, and cryptocurrencies to experiment and illustrates that the presence of alternative assets, such as cryptocurrencies, mean-variance methods underperform the benchmark portfolio among all different strategies.

This paper uses the experience of Alla Petukhina and Erin Sprunken (2021) for reference. It compares and evaluates various optimization techniques applied to a diversified investment universe in the following dimensions: several asset classes and a specific comparative view on the equally weighted portfolio.

The critical focus of the proposal is the application of different asset classes and the use of different kinds of optimization techniques to help increase return or decrease portfolio risk. In the empirical analysis part, we want to use some performance metrics to evaluate these strategies' performance from different aspects. And the main difference with Alla Petukhina and Erin Sprunken (2021) is that we consider the market expectation and contain the VIX INDEX in the portfolio and to see that whether there is a correlation between the VIX INDEX and other cryptocurrencies and also to find how the VIX INDEX influences on the investors' decision making when they are trying to create a market investment portfolio.

Our contribution is to compare which of the different strategic options can maximize the rate of return and whether there are any correlations and differences between containing VIX INDEX and the one without VIX INDEX in other investment portfolios.

The paper is structured as follows:

Section 2 contains experimental analysis data and explains the selection of the assets in the portfolio.

Section 3 explains the methodology of the investing strategies, containing the parameter estimation and the optimization calculations.

Section 4 provides a test and follow-up empirical analysis to evaluate differences amongst different strategies and check whether VIX INDEX positively impacts the portfolio's performance.

Section 5 gives a brief conclusion for the whole experiment and provides some feedback and reflection for the entire research.

**2 DATA**

There are three main classes of assets in our portfolio for the whole research: cryptocurrencies, gold, and VIX INDEX.

The data set of CCs contains the six largest by market capitalization as of 11.4.2021, obtained from investing.com. The daily observations start at 2018.10.31 up to 2021.10.31. In contrast to the gold, CCs also have comments on weekdays. The prices/exchange rates are measured per unit of cryptocurrency (i.e., a 6 USD price means that with 6 USD, one could buy 1 BTC). The trading volume is also measured in USD.

To be more specific, due to the data's availability, some central CCs contain many missing values; six different cryptocurrencies include BITCOIN, ETH, Binance Coin, Tether, Cardano, and XRP. You can find the data of the cryptocurrency in the appendix.

VIX INDEX is the Cboe Volatility Index, a real-time index representing the market's expectation for the relative strength of near-term price changes of the S&P 500 index because it is often seen as a way to gauge market sentiment and, in particular, the degree of fear among market participants. When VIX INDEX rises like the price of assets, investors are more likely to buy and hold more the investment of VIX INDEX and thus reduce the share of other assets in the portfolios. Besides, VIX INDEX is seen as an independent asset, adding to the portfolios. It could be seen as a method to avoid market risk. And in the empirical part, we want to explore its rationality and link to the other two classes of assets in the portfolio. Other possible influencing factors will also be analyzed and considered in this part, such as the investors' confidence, changes in legislation control conditions, Etc.

It is worth noting that even though the VIX index is not a specific kind of independent and tradable asset in the real market, the VIX index is calculated by taking a total of eight sequences of the call and putting options closest to parity in the recent and next months of the S&P500 index, respectively, and then weighted average their implied volatility. The concrete function of the VIX index is knowable. It means that we could use some methods and models, such as delta-vega neutral strategy, to devise a blanket of assets, including VIX futures, S&P 500 futures, S&P 500 options, and S&P 500 futures options to replicate the effectiveness of the tradable VIX index as a specific asset. [(2009)](2016)

Only VIX INDEX is not enough to hedge asset risks brought by cryptocurrencies in the short term. To bear risks in the long-term, we chose Gold as a risk-proof asset and added it to our investment portfolio. Gold is generally seen as a safe asset, and investors always buy it when they fear bear markets or crashes.(2010) In this way, no matter how the cryptocurrency market changes, we can maintain the value of the asset portfolio to the greatest extent and significantly reduce the risk.

Although gold has a value preservation effect in the case of long-term ownership, we cannot abandon VIX INDEX in our asset portfolio when we have gold. Because gold is too stable, when cryptocurrencies experience a cliff-like decline in a short period, the excellent stability prevents it from generating a corresponding proportional hedge against the erosion of cryptocurrencies in a short period. Because VIX INDEX is derived from SXP, its volatility and increase rate are similar to stocks, which is very high. It will also be better than gold to hedge the impact of cryptocurrencies in the short term. Therefore, VIX INDEX and gold are indispensable in our research. Moreover, the stability of gold can also avoid some unnecessary errors in the experiment. The short volatility allows gold to have no disproportionate impact in the case of calculation errors, and it can still prevent it to a certain extent—the role of risk.

Since the transaction of cryptocurrencies is going on every day, but gold is not, we did the data cleansing for the raw data. The raw data contain 1096 days initially, after removing the weekend days, and 90-days data is used for the moving window; the final data only includes 690 days. And the formula to obtain daily returns from prices is:

$$r_{i,t} = \frac{P_{i,t} - P_{(i,t-1)}}{P_{i,t-1}} = \frac{P_{i,t}}{P_{i,t-1}} - 1 \tag{1}$$

The index $i$ represents the sum of attributes and specific materials, and $t$ represents the corresponding time index. On that day, $P$ was related to price.

**3 Methodology**

In this section, the theoretical background and methodology are explained. It starts with a brief overview of the notation used, followed by the estimators for the necessary parameters. Afterward, a description of the strategies takes place. For each allocation rule, intuition, as well as the mathematical definition, is given. Lastly, the performance measurements used to evaluate the allocation techniques are Explained.

$T$ represents the number of available observations, equivalent to the total number of days in the data.

$N$ is the number of risky assets.

$u$ is a $N \times 1$ vector of expected returns of these assets and $r$ the actual, realized returns.

$\Sigma$ is the $N \times N$ variance-covariance matrix of the same risky assets.

$r_f$ is a scalar representing the risk-free rate.

$1_N$ represents a vector of ones of length $N$:$(1_1,1_2 1...,1_i,...,1_N)^T$.

$x$ is a $N \times 1$ vector of the weights: $(x_1, x_2, ..., x_i, ...,x_N)^T$

$M$ is a scalar representing the window size for moving-window estimations.

if $i = 8$, it means the 8th asset is VIX index; and if $i = 7$, it means the 7th asset is gold; and if $i = 1,2,3,4,5,6$, it means they are the main six kinds of CCs.

## 3.1 Parameter estimation

The parameters for the respective strategies are not known a priori. To implement them, estimators are necessary, which are described in this section. The window sizes is essential, as all parameters are estimated on a rolling-window basis. That means new information is continuously included in the parameters, and data points older than M are dropped out of the estimation. The abbreviation in the brackets will be used in the empirical section to denote which estimator was used. The estimators are dependent on time; that is, for every point of time $t$, a parameter is estimated based on the respective window.

### 3.1.1 Arithmetic mean (AM)

The first parameter refers to the unknown $\mu$, which represents a vector of expected Returns $(\mu_1,...,\mu_i,...,\mu_N)^T$ and has to be estimated for every time $t$. Here, this will be the arithmetic mean, described by the following formula:

$$\hat{u}_{i,t} = \frac{1}{M}\sum_{j=t-M}^{t-1} r_{i,j} \qquad (2)$$

There, i corresponds to asset i of the set, t to a time, r is the realized return.

### 3.1.2 Geometric mean (GM)

However, the arithmetic mean might not be suitable as an estimator for mean growth rates. By definition of Eq. (1) returns are growth rates of prices. Thus, we include the geometric mean as another estimator, especially as prices follow a geometric series. Furthermore, the geometric mean likely is a more conservative estimator and therefore, might lead to better results. The GM is traditional because, for positive real numbers, the GM is never more significant than the same sample's arithmetic mean. Furthermore, Jacquier et al. (2003) showed that compounding the arithmetic average is an upwardly biased estimator. The following formula represents the computation of the geometric mean:

$$\hat{u}_{i,j} = \sqrt[M]{\prod_{j=t-M}^{t-1}\left(1+r_{i,j}\right)} - 1 \qquad (3)$$

Again, $i$ corresponds to asset $i$ of the set, t to a time, $r$ is the realized return and $M$ the window sizes.

### 3.1.3 Variance–covariance matrix (AM/GM)

Besides the parameter $\mu$, the variance-covariance matrix $\Sigma$ is used. The usual estimator takes the following form:

$$\hat{e}_{ij,t} = \frac{1}{M-1}\sum_{h=t-M}^{t-1}\sum_{k=h-M}^{t-1}(r_{i,h} - \hat{u}_{i,t})(r_{j,k} - \hat{u}_{j,t}) \qquad (4)$$

$$\hat{\Sigma}_t^\mu = [\hat{e}_{ij,t}]. \qquad (5)$$

The respective window size M will always be the same as for the $\mu$ estimator here the "u" as an exponent is used to assign a name ("u" stands for usual).

**3.2 Portfolio optimization**

In this subsection, the different allocation strategies are discussed. For better readability, the time index has been omitted. However, it should be noted that the weights and necessary parameters are estimated for each rebalancing. Thus, $x_t$, $\mu_t$ and $\Sigma_t$ are everywhere in this subsection. In the empirical section, the rebalancing has been conducted daily. If not explicitly stated, short selling and leverage of CCs and gold are not allowed, but the VIX index allows for short selling. However, during the process of short-selling, the investor is not permitted to use leverage, which means he could not use the money from short-selling of the VIX index to buy the other seven assets. The weight of short selling could not exceed one, meaning that the market value of the VIX index. The investor borrows for short selling could not exceed the total value of his assets. This constraint condition prevents investors from lending money far more than their ability allows in some investment strategies to maximize their utility (as will be the case in the strategy regarding maximizing CRRA utility function later in part). It will result in excessive leverage and risk, which is unreasonable and dangerous in the market.

**3.2.1 Equally weighted**

The Equally weighted means that all the assets in the portfolio have the same weight. For example, if there are assets in the portfolio, asset A and asset B, each weighs 50% of the whole portfolio. Suppose the portfolio's weight is readjusted every day to hold 50% of each of the two assets. In that case, the daily return of the portfolio with the exact weight is only 50% of the return of investment A plus 50% of the return of asset B(daily). The naive rule has a few convenient features: it is easy to implement and non-parametric; it is diversified and characterized by low trading costs. It can be described as the formula shows:

$$x_i^{EW} = \frac{1}{N}, \forall_i \in [1,N]. \tag{6}$$

In this equation, $i$ is the index for the $i$ the asset.

This approach will serve as a benchmark in the empirical section.

### 3.2.2 Mean-variance (modern portfolio theory—MPT)

The MPT was one of the milestones in financial economics history, founded by the Nobel Prize laureate Harry Markowitz's work, see Markowitz (1952). Mean-variance (modern portfolio theory) assumes that all the investors are rational to make the investment decision if they have complete information. They constantly seek low risks and high returns. To introduce the mean-variance analysis, these two components, variance and expected return are the most important. A variance is a number that indicates how different or scattered the numbers are. For example, a spread can tell you how the return from particular security is distributed daily or weekly. The expected return is the probability that indicates the expected return on a securities investment. The investors always choose the one with a lower variance if the expected returns are the same. Similarly, they would like to select the higher expected return if the same variances. And Portfolios above the efficient frontier are not feasible; see Markowitz (1952) and Elton et al. (2003)

We aim to find the optimal strategy along the efficient frontier. It is also possible and correct if the frontier is a straight line. Portfolios can be below the limit but not optimal, as investors have at least the same income but pay less risk, or the same risk, but higher. Portfolios could not build beyond adequate limits indeed. And to approach the central idea, the target return can be formulated as:

$$\min_{x} x^T \Sigma x \tag{7}$$
$$s.t.\ x^T u \geq \mu^{target}$$
$$s.t.\ x^T 1_N \leq 1$$
$$s.t.\ \sum_{i=1}^{7} x_i \leq 1$$
$$s.t.\ x_i \in [0,1]$$
$$s.t.\ x_8 \in [-1,1]$$

### 3.2.3 Global Minimum-Variance

The minimum variance is one part under the Mean-Variance Model. However, this strategy is an exceptional case. It has only one input parameter-the mean-variance of the variance and ignoring funds for optimization is equivalent to assuming that all funds are equal, see DeMiguel et al. (2009)

$$\min_{x} x^T \Sigma x \tag{8}$$
$$s.t.\ x^T 1_N \leq 1$$
$$s.t.\ \sum_{i=1}^{7} x_i \leq 1$$
$$s.t.\ x_i \in [0,1]$$
$$s.t.\ x_8 \in [-1,1]$$

This approach is the riskiest of all Mean Variance strategies as it is the lowest point in the efficient frontier.

**3.2.4 Sharpe Ratio Maximization**

The Sharpe ratio Maximization, see Sharpe (1966), is another type of the Mean-Variance Model. This system will be optimized and measure the effectiveness of asset reports in proportion to the risk of rising returns. The explanation of this is:

$$\widehat{\Psi}_k^{sr} = \frac{u-r_f}{\sqrt{\hat{\sigma}^2}} = \frac{x^T u - r_f}{\sqrt{x^T \Sigma x}} \tag{9}$$

Reliable profitability data is evaluated based on historical data. Various asset classification models have always been used as risk sources. No risk return can be attributed to returns. Sharpe's high scores indicate a more risk-efficient portfolio as well. Thus, to build a portfolio on it, which is called the tangency portfolio, the expression is:

$$\max_x \frac{x^T(u-r_f)}{(x^T \Sigma x)^{\frac{1}{2}}} \tag{10}$$

$$s.t.\ x^T 1_N \leq 1$$

$$s.t.\ \sum_{i=1}^{7} x_i \leq 1$$

$$s.t.\ x_i \in [0,1]$$

$$s.t.\ x_8 \in [-1,1]$$

**3.2.5 Isoelastic Utility and CRRA (Coeffective of Relative Risk Aversion)**

In economics, the utility function is a quantitative description of preference, as we all know. The value of the utility function is consistent with the relationship between choices, which is the basis for investors' decision-making. Finance makes decisions in an uncertain environment. Financial asset price and investment return are random variables, so we must first determine their utility function.

Expected Utility was developed in the 1950s by Von Neumann and Morgenstern using logic and mathematical tools based on axiomatic assumptions. It defines the value of utility function for different possible results in the uncertain environment. It uses this expected utility function to describe the probability distribution of uncertainty and take the expected value.

Expected Utility Theory is defined that: If some random variable $X$ is evaluated with probability $P_i$ $x_i, i = 1,2... ,n$, and the utility $u(x_i)$ of a person when he is sure to obtain $x_i$, then the utility given to him by the random variable is:

$$U(X) = E[u(X)] = P_1 u(x_1) + P_2 u(x_2) + \ldots + P_n u(x_n) \tag{11}$$

E[u(X)] represents the expected utility of random variable X., So U of X is the desired utility function.

The investor wants to maximize his expected utility; thus, the standard utility maximization investor's problem can be formulated as:

$$\begin{aligned} \max E[u(r_{t+1})] &= E[u(x^T \mu_{t+1})] \\ s.t.\ x^T 1_N &\leq 1 \\ s.t.\ \sum_{i=1}^{7} x_i &\leq 1 \\ s.t.\ x_i &\in [0,1] \\ s.t.\ x_8 &\in [-1,1] \end{aligned} \tag{12}$$

The isoelastic utility function expresses utility in consumption or other economic variables important to decision-makers. The isoelastic utility function is a particular case of absolute hyperbolic risk aversion. It is also the only utility function with continuous relative risk aversion, so it is also called the CRRA (Coeffective of Relative Risk Aversion) utility function. The Investor has a CRRA (constant relative risk aversion) utility, as it is often assumed in the literature (A ıt-Sahalia and Brandt [2001](#)). The reason behind the use of this particular utility function is clearly stated by Brandt et al. ([2009](#)): "the advantage of CRRA utility is that it incorporates preferences toward higher moments in a parsimonious manner. In addition, the utility function is twice continuously differentiable, which allows us to use more efficient numerical optimization algorithms that make use of the analytic gradient and Hessian of the objective function". However, it is worth noticing that the proposed methodology applies to any other type of utility function. CRRA utility function is defined as:

$$u(r_{t+1}) \begin{cases} \frac{(1+r_{t+1})^{1-\gamma}}{1-\gamma} \; if \; \gamma \geq 0, \gamma \neq 1 \\ \ln(1+\gamma_{t+1}) \; if \; \gamma = 1 \end{cases} \tag{13}$$

Where $\gamma$ represents the relative risk aversion coefficient (the higher the value of $\gamma$, the more risk-averse the investor is). Combining Eq.10 with Eq.11, we set the relative risk aversion coefficient $\gamma \geq 0$, and thus, we could get the final objective function formula:

$$\max E[u(r_{t+1})] = E[u(x^T u_{t+1})] = \sum_{j=t-1}^{t-M} \frac{1}{M} \left[ \frac{(1+x^T r_j)^{1-\gamma}}{1-\gamma} \right] \tag{14}$$

$$s.t. \; x^T 1_N \leq 1$$

$$s.t. \; \sum_{i=1}^{7} x_i \leq 1$$

$$s.t. \; x_i \in [0,1]$$

$$s.t. \; x_8 \in [-1,1]$$

### 3.2.6 Adjusted Sharpe ratio

The sharpe ratio was already mentioned in Sect. 3.2.4. Economically, it can be interpreted as how much return an investor receives per unit of risk. Colloquially, it is how much return an investor could "buy" by paying an additional team of risk $\sigma$. Formally, it is defined as:

$$\widehat{\Psi}_k^{sr} = \frac{u - r_f}{\sqrt{\hat{\sigma}^2}} = \frac{x^T u - r_f}{\sqrt{x^T \Sigma x}} \tag{15}$$

The quick ratio can be interpreted as how much return an investor receives per unit of risk. Colloquially, it is how much return an investor could "buy" by paying an additional team of risk $\sigma$. However, investors might be interested in skewness and kurtosis as well; thus, to assess the performance properly, Pézier and White (2008) proposed the adjusted sharpe ratio:

$$\widehat{\Psi}_k^{asr} = \left[1 + \left(\frac{\hat{S}}{6}\right) \Psi_k^{sr} - \left(\frac{\hat{K}}{24}\right) \left(\widehat{\Psi}_k^{sr}\right)^2\right] \tag{16}$$

This formula, Ŝ and K̂ represent sample skewness and excess kurtosis in this formula. This measurement incorporates the preference for positive skewness and excess kurtosis, penalizing the opposite. This is important, as a distribution with negative skewness and positive excess kurtosis increases the tail risks. Investors do not desire these. Therefore, the objective function formula is:

$$\max \widehat{\Psi}_k^{asr} = \left[1 + \left(\frac{\hat{S}}{6}\right) \Psi_k^{sr} - \left(\frac{\hat{K}}{24}\right) \left(\widehat{\Psi}_k^{sr}\right)^2\right] \tag{17}$$
$$s.t. \ x^T 1_N \leq 1$$
$$s.t. \ \sum_{i=1}^{7} x_i \leq 1$$
$$s.t. \ x_i \in [0,1]$$
$$s.t. \ x_8 \in [-1,1]$$

Skewness is defined as:

$$\hat{S} = E\left[\left(\frac{x^T r - \mu}{\sigma}\right)^3\right] = \frac{1}{\sigma^3} E[(x^T r - x^T u)^3] = \frac{1}{\sigma^3} \sum_{j=t-1}^{t-M} \frac{1}{M} [x^T(r_j - u)]^3 \quad (18)$$

Coskewness is defined as:

$$s_{ijk} = E\{[r_i - E(r_i)][r_j - E(r_j)][r_k - E(r_k)]\} = \frac{1}{M} \sum_{l=t-1}^{t-M} (r_{i,l} - u_i)(r_{j,l} - u_j)(r_{k,l} - u_k) \quad (19)$$

According to Eq.16 and Eq.17:

$$\hat{S} = \frac{1}{\sigma^3}\left[\sum_{i=1}^{N} x_i^3 s_{iii}^3 + 3\sum_{i=1}^{N}\left(\sum_{j=1\, i\neq j}^{N} x_i^2 x_j s_{iij} + \sum_{j=1\, i\neq j}^{N} x_i s_{ijj}\right) + \sum_{i=1}^{N}\sum_{j=1\, j\neq i}^{N}\sum_{k=1\, k\neq i\, k\neq j}^{N} x_i x_j x_k s_{ijk}\right] = \frac{1}{\sigma^3}\sum_{i=1}^{N}\sum_{j=1}^{N}\sum_{k=1}^{N} x_i x_j x_k s_{ijk} \quad (20)$$

Therefore, the co-skewness matrix $RM_3$ can be computed as a $N \times N^2$ matrix. (2004)(2017) According to this procedure, the daily portfolio skewness can be computed as

$$\hat{S} = x^T RM_3 (x \otimes x) \quad (21)$$

where $RM_3$ is the co-skewness matrix and $\otimes$ represents the Kronecker product. The co-skewness matrix corresponds to $N$ matrixes $A_{ijl}$ of dimension $N \times N$ such that

$$RM_3 = [A_{1jl}\ A_{2jl}\ A_{3jl}...\ A_{Njl}] \quad (22)$$

Where each element, $a_{ijl}$, is given by

$$a_{ijl} = s_{ijl} \tag{23}$$

Analogous to the skewness approach,

$$\widehat{K} = E\left[\left(\frac{x^T r - \mu}{\sigma}\right)^4\right] = \frac{1}{\sigma^4}\sum_{i=1}^{N}\sum_{j=1}^{N}\sum_{l=1}^{N}\sum_{m=1}^{N} x_i x_j x_l x_m k_{ijlm} \tag{24}$$

Coskewness is defined as:

$$k_{ijlm} = E\{[r_i - E(r_i)][r_j - E(r_j)][r_l - E(r_l)][r_m - E(r_m)]\} = \frac{1}{M}\sum_{h=t-1}^{t-M}(r_{i,h} - u_i)(r_{j,h} - u_j)(r_{l,h} - u_l)(r_{m,h} - u_m) \tag{25}$$

The daily portfolio kurtosis can be obtained by computing the following products

$$\widehat{K} = x^T RM_4 (x \otimes x \otimes x) \tag{26}$$

where $RM_4$ represents the cokurtosis matrix. The $RM_4$ matrix corresponds to $N^2$ matrixes $B_{ijlm}$ of dimension $N \times N$ such that (2015)

$$RM_4 = [B_{11lm}\ B_{12m}\cdots B_{1Nlm}\ B_{21lm}\ B_{22lm}\cdots B_{2Nlm}\cdots B_{N1lm}\ B_{N2lm}\cdots B_{NNlm}] \tag{27}$$

where each element, $b_{ijlm}$, is given by

$$b_{ijlm} = k_{ijlm} \tag{28}$$

### 3.3 Performance metrics

To assess and compare the performance of the different strategies, it is necessary to introduce comparable evaluation metrics. Let $\widehat{\Psi}$ denote the estimator for success measurement.

$\hat{\mu}_k$ and $\hat{\sigma}_k^2$ denote the arithmetic mean return and variance of the respective kth strategy.

### 3.3.1 Turnover

An important dimension to assess the performance of a portfolio is the number of trading fees necessary to implement a strategy. We use a turnover to proxy transactional costs of strategy with the following computation:

$$\Psi_k^{to} = \frac{1}{T-M}\sum_{t=M+1}^{T}\sum_{i=1}^{N}\left(|\hat{x}_{k,i,t+1} - \hat{x}_{k,i,t}|\right) \qquad (29)$$

Here, $\hat{x}$ denotes the weight on asset i, in time t + 1 after rebalancing and right before rebalancing (t+). k denotes the k-th strategy. It can be seen that this formula calculates the absolute sum of changes in the weights, so the interpretation is that the more significant this metric is, the higher the implementation cost of the strategy.

### 3.3.2 Terminal return

The last metric used is the terminal return, sometimes called terminal wealth. It is an essential factor because it denotes a strategy's outcome at the end of the investing period. It does not control for risk in any way. This metric's importance lies in the fact that a fund manager's performance may be measured by the wealth she created for her investors. The following formula is used for the computation:

$$\Psi_k^{tr} = \prod_{t=M+1}^{T}(1 + r_{t,k}) \qquad (30)$$

Here, r denotes the realized portfolio return, k the k-strategy, and t is the time index. M is the window size. This formula, thus, represents cumulative performance at the final time T and can also be used to calculate the daily compound return.

### 3.3.3 Sortino ratio

The Sortino ratio is a measure of the relative performance of a portfolio. It is like the Sharpe ratio, but the Sortino ratio uses the downward biased standard deviation rather than the total standard deviation to distinguish adverse and favorable fluctuations. Like the Sharpe ratio, the higher this ratio indicates that the fund can obtain a higher excess return by taking the downside risk of the same unit.

$$\Psi_k^{str} = \frac{\hat{\mu}_k - r_f}{\sigma_k} \qquad (31)$$

$\sigma_k$ is the downlink standard deviation

### 3.3.4 Certainty equivalent

The certainty equivalent gives information about the rate a risk-free asset must return at least, such that an investor was indifferent between the respective kth portfolio and the risk-free investment.

The certainty equivalent takes the form:

$$\widehat{\Psi}_k^{ceq} = \hat{u}_k - \frac{\gamma}{2}\hat{\sigma}_k^2 \qquad (32)$$

### 4 Empirical Analysis

We evaluate and discuss the empirical results in this section. The entire unit is divided into three parts. The first part is the horizontal comparison of the results of the same portfolio under different strategies. The second part compares having VIX INDEX and no VIX INDEX in the portfolio under the same method. And the third part is the comparison between the result of having VIX INDEX and no VIX INDEX in the portfolio when the cryptocurrency market is not good. We employed two different parameters for all parametric calculations. To express more conveniently, we use the abbreviations in the parameters, where (AM) is Arithmetic Mean and (GM) is the Geometric Mean.

| Parameter | Description | Value |
|---|---|---|
| $r_f$ | Risk-free Rate | 0 |
| $u^{target}$ | Target Return (MPT) | $2.6 \times 10 \times (-4)$ |
| M | Window Size | 90 |
| $W_0$ | Starting Portfolio Value at 0 | 1 |
| $\gamma$ | Risk Aversion | 5 |

First, values for the other parameters explained in the methodological section need to be assigned (see Table 1).

|  | CEQ | TO | S | TR |
|---|---|---|---|---|
| Equally Weight | 0.003523879 | 0 | 0.154027868 | 8.281734495 |
| Global Minimum – AM | 8.32196E-05 | 0.007188916 | 0.05167959 | 1.057658295 |
| Global Minimum – GM | 8.30944E-05 | 0.007225653 | 0.051674303 | 1.057570361 |
| Target – AM | 0.000389053 | 0.252334243 | 0.073033352 | 1.288554187 |
| Target – GM | 0.000883243 | 0.069664682 | 0.342661418 | 1.767177486 |
| ASR – AM | 0.004402771 | 0.65447421 | 0.166727428 | 13.2360353 |
| ASR – GM | 0.004747624 | 0.661963017 | 0.172547537 | 16.41627608 |
| SR – AM | 0.004355 | 0.639147 | 0.190981 | 13.46407 |
| SR – GM | 0.005221 | 0.662397 | 0.183507 | 22.1755 |
| CRRA | 0.033744127 | 1.364666701 | 2.878333703 | 4300521480 |

The chosen target return might seem very low, but the observations and the calculated (expected) returns are daily. The denoted return corresponds to an approximate return of 10% per year. Furthermore, the assumption of $r_f = 0$ implies that returns and excess returns are equal.

The risk aversion $\gamma$ of the investors is uniformly assumed as five. And the rolling window size (M) we set as 90 days.

**4.1 Same portfolio under different strategies**

There are ten sets of different strategies with AM and GM parameters. We choose equally weighted as the benchmark and compare it with other strategies, as it is the most naive and straightforward strategy.

Table 1 illustrates that the Global Minimum strategy has the highest CEQ among all strategies. CEQ is the certainty equivalent representing the amount of guaranteed money that investors would accept now instead of taking the risk of getting more money at a future date. A high CEQ reveals that the risk premium is meagre under the same expected return. Since the higher the risk premium, the higher the risk the investors undertake. The Global Minimum strategy with the highest CEQ is the most stable, and it is the best one for the risk-averse investors amongst all strategies we focus on. CRRA strategy only focuses on the terminal return, which is why CRRA owns the highest terminal return among all strategies. However, some potential risks need to be considered.

For the two strategies of SR and ASR, we should compare and discuss the two aspects of skewness and kurtosis. The two diagrams below are the return density images of the SR strategy and ASR strategy with AM as the parameter and the SR strategy and ASR strategy with GM. SR is green, and ASR is red. First, let's look at the Kurtosis part. In the image with AM as the parameter, the kurtosis of SR is 2.1790468, and the kurtosis of ASR is 3.64304923. So, SR is platykurtic with a thinner tail, and ASR is leptokurtic with a fatter tail. A fatter tail means there is a higher probability of having extreme value. So, the thinner-tailed SR has a better performance than ASR. When the parameter is GM in this set of data, the kurtosis of SR is 3.356380809, and the kurtosis of ASR is 4.6628285985. The comparison of this data set is the same as the comparison of the previous data collection. SR always has lower kurtosis. So, from the perspective of kurtosis, SR is a better strategy because it has a thinner tail and lower risk. From the standpoint of skewness, investors prefer to be negatively skewed because such skewness will make most returns higher than the mean return, and the rate of return is generally higher. Of course, right-skewed distribution has the opposite effect. When the parameter is AM, the Skewness of SR is -0.054671023, and the skewness of ASR is 0.6324623898. It can conclude that the skewness of SR is smaller than ASR, so it is a left-skewed distribution, the mode is higher than the mean, so the terminal return of SR is often higher than ASR. When GM is used as a parameter, the skewness of SR is 0.69308987, and the skewness of ASR is 0.9163852885. The conclusion of this data is roughly the same as that of the previous AM group. SR still has a negatively skewed distribution and a relatively higher level of terminal return. In the part of skewness, SR also has better performance than ASR.

A negatively skewed distribution may indeed lead to a very negative expectation value and the possibility of a considerable loss. Because the kurtosis of SR is smaller than ASR, the dispersion of returns is smaller, so the possibility of highly negative expectations is greatly reduced. The benefits brought by SR can make us temporarily ignore his case the potential risks brought by contacting us to analyse the kurtosis part. Thus, SR still has better performance than ASR.

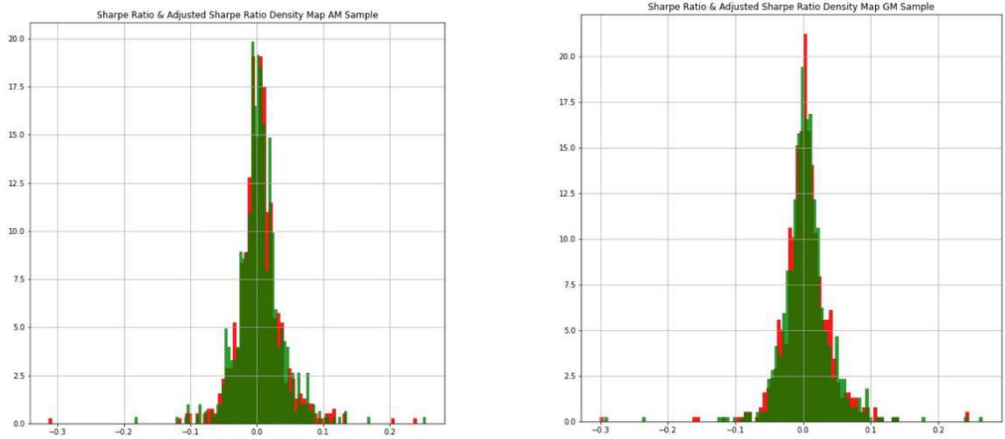

## 4.2 Same strategies with or without VIX INDEX

In this part, we use two portfolios, the one with VIX INDEX and the other one without VIX INDEX, in the same strategies. Because of the vertical analysis, we only focus on the terminal return in each strategy to compare and seek if the VIX INDEX has a positive influence on our portfolio.

| VIX | GM-AM | GM-GM | ASR-AM | ASR-GM | T-AM | T-GM | CRRA | EW | SR-AM | SR-GM |
|---|---|---|---|---|---|---|---|---|---|---|
| TO | 0.006285495 | 0.006294057 | 0.590782798 | 0.60203041 | 0.20966889 | 0.058227924 | 1.301705263 | 0 | 0.564815 | 0.609174 |
| CEQ | 5.95237E-05 | 5.97349E-05 | 0.003224757 | 0.005056383 | 0.000240512 | 0.000343427 | 0.031510626 | 0.003639331 | 0.003588 | 0.004946 |
| S | 0.037426648 | 0.037604424 | 0.108550807 | 0.200841657 | 0.091216562 | 0.164160801 | 2.108536723 | 0.127913789 | 0.125187 | 0.196846 |
| TR | 1.04052601 | 1.040680691 | 5.788150111 | 21.04609663 | 1.174657866 | 1.263380593 | 1022271021 | 7.820818972 | 7.48363 | 19.64142 |

| VIX | GM-AM | GM-GM | ASR-AM | ASR-GM | T-AM | T-GM | CRRA | EW | SR-AM | SR-GM |
|---|---|---|---|---|---|---|---|---|---|---|
| TO | 0.007188916 | 0.007225653 | 0.65447421 | 0.661963017 | 0.252334243 | 0.069664682 | 1.364666701 | 0 | 0.639147 | 0.662397 |
| CEQ | 8.32196E-05 | 8.30944E-05 | 0.004402771 | 0.004747624 | 0.000389053 | 0.000883243 | 0.033744127 | 0.003523879 | 0.004355 | 0.005221 |
| S | 0.05167959 | 0.051674303 | 0.166727428 | 0.172547537 | 0.073033352 | 0.342661418 | 2.878333703 | 0.154027868 | 0.190981 | 0.183507 |
| TR | 1.057658295 | 1.057570361 | 13.2360353 | 16.41627608 | 1.288554187 | 1.767177486 | 4300521480 | 8.281734495 | 13.46407 | 22.1755 |

Comparing these two data sets, it is not difficult to see those portfolios with VIX INDEX generally have higher terminal returns. From the data point of view, VIX INDEX has indeed played a positive role in the portfolio. Fundamentally, we have also done covariance of various assets and terminal returns.

|      | GM-AM        | GM-GM        | ASR-AM   | ASR-GM       | T-AM     | T-GM     | CRRA     | EW | SR-AM    | SR-GM       |
|------|--------------|--------------|----------|--------------|----------|----------|----------|----|----------|-------------|
| BTC  | -5.58074E-08 | -6.61E-08    | -0.00031 | -0.00038     | 2.78E-06 | -2.8E-06 | -0.00025 | 0  | -0.00033 | -0.00037    |
| ETH  | -1.64307E-07 | -1.6959E-07  | -0.00012 | -7.32567E-05 | 0.000147 | 1.56E-05 | -0.0003  | 0  | -5.9E-05 | -0.00012    |
| XRP  | -5.4742E-07  | -5.69571E-07 | -5.6E-05 | 1.32815E-05  | 4.35E-05 | 1.6E-05  | -0.00025 | 0  | -4.7E-05 | -3.2E-05    |
| LTC  | -8.33072E-05 | -8.33637E-05 | -9.8E-05 | -0.000177295 | 0.003524 | 0.000607 | 0.000221 | 0  | -0.0001  | -8.9E-05    |
| XMR  | -3.60989E-07 | -3.72616E-07 | -7.3E-05 | -5.883E-05   | 4.3E-06  | -6.5E-06 | -0.00018 | 0  | -9.8E-05 | -9.1E-05    |
| XLM  | -3.69612E-07 | -3.767E-07   | -0.00032 | -0.000360546 | 6.29E-05 | 8.24E-05 | -9.5E-05 | 0  | -0.00046 | -0.00039    |
| VIX INDEX | 1.02829E-05 | 1.02888E-05 | 5.36E-05 | 5.43166E-05 | 0.00295 | 0.002178 | 4.95E-05 | 0 | 6.55E-05 | 5.24421E-05 |
| GOLD | 7.87829E-05  | 7.89023E-05  | 0.000933 | 0.001001889  | -0.0041  | -0.0024  | 0.000827 | 0  | 0.001051 | 0.001053374 |

The data points out that VIX INDEX and gold have positive covariance with TR, but almost all cryptocurrencies have a negative covariance with TR. A positive covariance means that asset returns move together, while a negative covariance means they move inversely. So higher VIX INDEX and GOLD lead to a higher terminal return.

In addition, we also calculated the correlation between each asset and VIX INDEX.

|      | GM-AM        | GM-GM        | ASR-AM   | ASR-GM    | T-AM     | T-GM     | CRRA     | EW | SR-AM    | SR-GM       |
|------|--------------|--------------|----------|-----------|----------|----------|----------|----|----------|-------------|
| BTC  | -0.017515929 | -0.020854407 | -0.2304  | -0.28664  | 0.011563 | -0.0125  | -0.16335 |    | -0.22756 | -0.28864057 |
| ETH  | -0.074277934 | -0.076735293 | -0.08257 | -0.05236  | 0.176072 | 0.032161 | -0.19585 |    | -0.03937 | -0.08730129 |
| XRP  | -0.138640027 | -0.14470728  | -0.06054 | 0.013968  | 0.102077 | 0.040391 | -0.19167 |    | -0.04644 | -0.03488065 |
| LTC  | -0.185721812 | -0.185741539 | -0.1109  | -0.19825  | 0.283961 | 0.060263 | 0.179129 |    | -0.10297 | -0.09729476 |
| XMR  | -0.103310641 | -0.106675013 | -0.1011  | -0.07859  | 0.025764 | -0.06774 | -0.14202 |    | -0.11803 | -0.11388501 |
| XLM  | -0.059011815 | -0.060110456 | -0.22249 | -0.24299  | 0.094163 | 0.084532 | -0.07495 |    | -0.27946 | -0.26748463 |
| VIX INDEX | 1       | 1            | 1        | 1         | 1        | 1        | 1        |    | 1        | 1           |
| GOLD | 0.17508302   | 0.175261173  | 0.592172 | 0.619632  | -0.32121 | -0.23302 | 0.538568 |    | 0.591738 | 0.665196562 |

The correlation data shows that gold is positively correlated with VIX INDEX, but almost all cryptocurrencies are negatively correlated. The result shows that the growth of the VIX INDEX will also lead to the development of gold and the negative growth of cryptocurrencies. It can also prove that VIX INDEX can effectively impact the risks brought by cryptocurrencies, and both VIX INDEX and gold are significant in this portfolio. In this case,

**4.3 Same strategies with or without VIX INDEX in a bear market**

To explore whether VIX INDEX contributes to the portfolio's positive performance under any circumstances, we selected data from the 90-day cryptocurrency market downturn and re-calculated the estimated portfolio with VIX INDEX and the portfolio without VIX INDEX, and once again make a vertical comparison.

| VIX | GM-AM | GM-GM | ASR-AM | ASR-GM | T-AM | T-GM | CRRA | EW | SR-AM | SR-GM |
|---|---|---|---|---|---|---|---|---|---|---|
| TO | 0.001586324 | 0.001584104 | 0.252502526 | 0.258140711 | 0.01696878 | 0.005624694 | 0.521543667 | 0 | 0.246441 | 0.249778 |
| CEQ | 1.06457E-05 | 8.83926E-06 | 0.003375442 | 0.003375442 | -0.00022759 | 1.68908E-05 | 0.053691712 | 0.00444665 | 0.001 | 0.008341 |
| S | 0.003210405 | 0.002666365 | 0.09903108 | 0.20164784 | -0.05449613 | 0.004066785 | 3.17204396 | 0.146656381 | 0.028723 | 0.236499 |
| TR | 0.999147924 | 0.998653232 | 1.656567423 | 4.688910412 | 0.933366917 | 0.999831122 | 958956.5864 | 2.499892941 | 0.909324 | 6.354991 |

| NO | GM-AM | GM-GM | ASR-AM | ASR-GM | T-AM | T-GM | CRRA | EW | SR-AM | SR-GM |
|---|---|---|---|---|---|---|---|---|---|---|
| TO | 0.00118492 | 0.001181893 | 0.228185378 | 0.220141018 | 0.005848038 | 0.003923037 | 0.528546111 | 0 | 0.213644 | 0.227209 |
| CEQ | -2.4295E-05 | -2.1949E-05 | 0.003332235 | 0.00572371 | 1.92687E-05 | 2.95144E-05 | 0.050368167 | 0.004698148 | 0.003712 | 0.007259 |
| S | -0.00711832 | -0.00643266 | 0.10410978 | 0.15281324 | 0.005615753 | 0.007906545 | 2.789191963 | 0.132617706 | 0.105588 | 0.207513 |
| TR | 0.989324291 | 0.989977682 | 1.690912069 | 3.048626888 | 1.001376277 | 1.003958363 | 411250.9586 | 2.421417065 | 1.797504 | 4.768357 |

In this vertical comparison, we still focus on the Terminal Return (TR) only to compare the effect of VIX INDEX. By contrast, it is not difficult to see that, except for the four estimation methods of ASR-AM, T-AM, G-AM, and SR-AM, the others match the results of 4.2, that is, the portfolio with VIX INDEX has a higher terminal return and better performance than the portfolio without VIX INDEX. And in the estimation of these four abnormal results, the difference is very subtle—the negative impact caused by whether including the VIX index is not very significant. Therefore, the conclusion of 4.2 is still valid in this part. It is worth noting that VIX INDEX's positive effect on the cryptocurrency market's downturn may not be very significant, so investors should use VIX INDEX as a signal before making an investment decision, and it may be more effective than adding such assets directly to the portfolio. Of course, there is nothing wrong with joining the portfolio as well.

**5 Conclusion**

After the entire experiment has finished, we can draw some conclusions to help us better understand the relationship between cryptocurrency, VIX INDEX, and Gold. But of course, in many aspects, the design of this experiment may not be particularly perfect; there will be some loopholes are waiting to be perfected.

**5.1 Results**

Through the horizontal comparison of 4.1, we found that GM is relatively stable among all prediction models. CRRA is the best model to expand TR, but potential risks exist. The two indicators of ASR and SR give negative feedback, but to avoid the emergence of extreme values, the estimation of SR is relatively more reliable. And we have also explored the covariance of the three types of assets and TR and the correlation between cryptocurrency and gold and VIX INDEX, and we can draw conclusions that VIX INDEX is negatively correlated with the cryptocurrencies but positively correlated with gold. So VIX INDEX is an influential asset in the portfolio to hedge the risk from the cryptocurrency. That also shows that the VIX INDEX portfolio has a better performance than the portfolio without VIX INDEX. Analysis 4.3 can also reflect that VIX INDEX has positively influenced in general, no matter when the cryptocurrency market is rising or falling. The negative impact is so weak that we can temporarily ignore it. VIX INDEX can effectively help investors make better investment decisions as a signal that reflects market panic.

**5.2 Reflections**

We have some reflections and found some small loopholes that need improvement. I hope that researchers interested in this topic or related topics can improve and expand based on our experiments.

The first one originates from the raw data. Our data time is short, which is not long enough as a business cycle period. That is because the development time of cryptocurrencies is short, and we have no other solutions to deal with this defect. Thus, the performance of cryptocurrencies and VIX INDEX in each economic cycle is unknown, so it is ambiguous whether our conclusion is 100% correct.

The second concern is about the asset categories. There are only three categories of assets and eight assets total in our portfolio. The limit of the classes, especially six major cryptocurrencies, cannot fully represent the normal state of the cryptocurrency market, which causes many deviations in our results.

The third one is relative to the trading frequency. Our trading frequency is daily. For the T+0 trading market, high-frequency trading strategies may be more advantageous in this market, so the feasibility of high-frequency trading strategies for cryptocurrencies in the future needs to examine.

These three points are some reflections after we finished the experiment. There are some improvements in the follow-up. We can solve some, but we also did not consider some during the investigation.

## 5.3 Further Research

The technological means are still improving, and the cryptocurrency market is gradually improving and controlling. We believe that soon, the cryptocurrency market will continue to thrive. Researchers interested in the subject can also enrich the asset portfolio to make the entire experiment more complete and not prone to bias. At the same time, we also found that it is a good way if we can well employ the neural network in the AI field in the estimation of the asset portfolio, which will bring us a more accurate and precise estimation result, which can help us make better investment strategies in the field of cryptocurrency.